\documentclass[
    ,final            
  ]
  {aipproc}

\layoutstyle{6x9}

\usepackage{amsmath,amssymb,bm}

\newcommand{\nslash}{n \hspace{-0.21cm} / \;\;}

\begin{document}

\begin{flushright}
JLAB-TH-09-989
\end{flushright}

\title{What can we learn from the breaking of the Wandzura--Wilczek relation?
\footnote{Talk given by Alberto Accardi at ``Spin structure at long
  distances'', Jefferson Lab, March 12-13 2009.}
}

\classification{12.38.Bx, 13.60.Hb}
\keywords      {polarized DIS, higher twist,
  transverse polarization}

\author{Alberto~Accardi$^{a,b}$, Alessandro~Bacchetta$^{b}$,
  Marc~Schlegel$^{b}$}
  {address={
  $^a$Hampton University, Hampton, VA 23668, USA \\
  $^b$Jefferson Lab, Newport News, VA 23606, USA
}}


\begin{abstract}
We review the study the Wandzura--Wilczek relation for the structure function
$g_2$, with a particular attention on the connection with 
the framework of Transverse Momentum Dependent factorization. We emphasize that
the relation is broken by two distinct twist-3 terms.
In the light of these findings, we clarify what can be
deduced from the available experimental data on $g_2$, which indicate
a breaking of the order 20--40\%, and how to
individually measure the twist-3 terms. 
\end{abstract}

\maketitle


At large virtuality $Q^2$, the lepton-nucleon Deep Inelastic
Scattering (DIS) cross 
section scales with $x=Q^2/(2M\nu)$ modulo logarithmic
corrections, where $M$ is the target's mass and $\nu$ the virtual photon
energy. At low $Q^2$, power suppressed contributions become
important, e.g., target mass
corrections of $O(M^2/Q^2)$ and jet mass corrections of order
$O(m_j^2/Q^2)$ with $m_j$ the invariant mass of the current jet
\cite{Accardi:2008ne,Accardi:2008pc}, and  higher-twist (HT) corrections
of $O(\Lambda_{QCD}^2/Q^2)$ related to quark-gluon correlations inside
the nucleon \cite{Ellis:1982cd}. There are many reasons why we need to 
identify and measure higher-twist terms in
experimental data, for example,
(i) to verify quark-hadron duality \cite{Melnitchouk:2005zr},
(ii) to measure twist-2 Parton Distribution Functions (PDF) at large
fractional momentum $x$ and low-$Q^2$, e.g., the $d/u$ and $\Delta d/
d$ quark ratios, sensitive to the nonperturbative structure of the
nucleon\cite{Farrar:1975yb,Brodsky:1994kg,Isgur:1998yb},
(iii) to measure multiparton correlations, important to understand the
nucleon structure beyond the PDFs\cite{Burkardt:2009},
(iv) to determine the perturbative QCD evolution of the $g_2$ polarized
structure function among others\cite{Belitsky:1997ay},
(v) to calculate the high-$k_T$ tails of transverse momentum
dependent (TMD) parton distribution functions\cite{Ji:2006ub}. 
 
The inclusive DIS cross section is determined by the hadronic tensor
$W^{\mu\nu}$, defined as the imaginary part of the forward virtual
photon Compton scattering amplitude. $W^{\mu\nu}$ can be decomposed in 2
unpolarized structure functions, $F_{1,2}$, which we do not discuss
here, and 2 polarized structure functions $g_{1,2}$. Its antisymmetric
part reads
\begin{align}
W^{\mu\nu}(P,q) 
  = \frac{1}{P\cdot q} \varepsilon^{\mu\nu\rho\sigma} q_\rho 
  \Big[ S_\sigma g_1(x,Q^2) 
  + \Big( S_\sigma - \frac{S\cdot q}{P\cdot q} p_\sigma \Big) g_2(x,Q^2)
  \Big] \ ,
\end{align}
where $P,S$ are the target momentum and spin. Among the structure
functions, $g_2$ is unique because it is the only one with twist-3
contributions that can be measured in inclusive DIS. Furthermore, its
higher-twist contribution can be isolated thanks to the
Wandzura-Wilczek (WW) relation, first obtained in the Operator Product
Expansion (OPE) formalism \cite{Wandzura:1977qf}:
\begin{align}
  g_2(x,Q^2) = g_2^{WW}(x,Q^2) + \Delta(x,Q^2) \ .
\label{eq:WWrel}
\end{align}
Here $g_2^{WW}$ is determined by the leading twist (LT) part of
$g_1$, which is rather well known experimentally:
\begin{align}
  g_2^{WW}(x,Q^2) = -g_1^{LT}(x,Q^2) + \int_{x}^1 \frac{dy}{y}
  g_1^{LT}(y,Q^2) \ .
\end{align}
Strictly speaking, the WW relation is the LT part of \eqref{eq:WWrel}. 
The breaking term $\Delta$ is a pure HT term, meaning that its moments
are matrix elements of local operators of twist-3 or higher, with ``twist''
defined as dimension minus spin of the local operator
\cite{Jaffe:1989xx,Jaffe:1990qh}. In this talk we will limit our analysis to
twist-3 operators, and will drop the dependence on $Q^2$ for ease of
notation. Details can be found in Ref.~\cite{Accardi:2009}. 

\section{Parton distributions in perturbative QCD}

In perturbative QCD the structure functions can be expressed as a
convolution of a perturbatively calculable coefficient, and a number
of nonperturbative LT parton distributions and HT parton correlations.
In particular, the WW relation can be obtained in the framework of
collinear factorization \cite{Belitsky:1997ay} or in transverse
momentum dependent factorization, as we will shortly describe.

Let's define the quark-quark correlator
\begin{align}
  \Phi^a_{ij} (x,\vec{k}_T)
    = \int \frac{d \xi^- d^2\xi_T}{(2 \pi)^3} \, e^{i k \cdot \xi} 
    \langle P,S \, | \, \bar{\psi}_j^a(0) \, W(0,\xi| n_-) \, 
    \psi_i^a(\xi) \, | \, P,S \rangle\Big|_{\xi^+=0}
\end{align}
where $i,j$ are Dirac indices, $a$ is the quark flavor index,
$k$ its 4-momentum, $x = k\cdot n_- / P\cdot n_-$ its
fractional momentum and $\vec{k}_T$ its transverse momentum relative
to the parent nucleon. 
The ``plus'' and ``minus'' components of a 4-vector are defined as
$a^\pm = a \cdot n_\mp$ in terms of two orthogonal light-cone vectors
$n_+^2=n_-^2=0$ such that $n_-\cdot n_+ = 1$, and $n_+^\mu$ is
proportional to $P^{\mu}$ up to mass corrections. 
$W$ is a Wilson line (gauge link) whose precise
form depends on the process. The direction of the Wilson line is
determined by an additional 4-vector beside $P,S$, which in tree-level
analyses such as we pursue here is identified with the light-cone
vector $n_-$. In the light-cone gauge $n_-\cdot A=0$ the Wilson line
is identically equal to 1 and
\begin{align}\
  \Phi^a_{ij}(x,\vec{k}_T)\stackrel{\text{LC}}{=}
    \int \frac{d \xi^- d^2\xi_T}{(2 \pi)^3} \, e^{i k \cdot \xi} 
    \langle P,S \,|\, \bar{\psi}_j^a(0) \psi_i^a(\xi) \,|\, P,S 
    \rangle\Big|_{\xi^+=0} \ .
\end{align}
Nonetheless, the dependence on $n_-$ appears explicitly in the 
gauge field propagators and cannot be in general neglected.

For any Dirac matrix $\Gamma$ we define the projection 
$\Phi^{a[\Gamma]} = \text{Tr} [\Gamma \Phi^a]/2$. 
The relevant TMDs are defined as follows:
\begin{align*} 
  \Phi^{a[\gamma^+ \gamma_5]}(x,\vec{k}_T) & = S_L \, g_{1}^a(x,\vec{k}_{T}^{2}) 
    + \frac{\vec{k}_{T} \cdot \vec{S}_{T}}{M} \, g_{1T}^a(x,\vec{k}_{T}^{2}) 
\\ 
  \Phi^{a[\gamma^{i}\gamma_5]} (x,\vec{k}_T)& = \frac{M}{P^+} 
 S_{T}^{i} \, g_{T}^a(x,\vec{k}_{T}^{2}) 
    + \ldots 
\end{align*}
The inclusive DIS is determined by collinear parton distribution functions
(PDFs) which are defined by transverse momentum integration of the TMDs: 
$g_\sharp(x) = \int d^2k_T\, g_\sharp(x,\vec{k}_T)$ and 
$g_\sharp^{(1)}(x) = \int d^2k_T\, \frac{\vec k_T^2}{2M}
g_\sharp(x,\vec{k}_T)$, with $\sharp$ indicating any of the above defined
TMDs.

\section{Equations of motion and Lorentz invariance}

The Dirac equations of motions for the quarks, and the Lorentz
invariance of the theory imply the following 2 relations between
twist-2 and pure twist-3 functions:
\begin{align}
  \text{(EOM)} \hspace*{-1.5cm} &&  g_{1T}^{a(1)}(x) 
    & = x g_T^a(x) - x \tilde{g}^a_T(x) + O(m/M)
    \label{eq:EOM}\\
  \text{(LIR)} \hspace*{-1.5cm} &&  g_{T}^a(x) 
    & = g_{1}^a(x)+\frac{d}{dx}\,g_{1T}^{a(1)}(x) + \hat{g}_T^a(x)
    \label{eq:LIR}
\end{align}
where for light quarks we can neglect the term proportional to the
quark mass $m$ compared to a typical hadronic scale $M$.
$\tilde g$ and $\hat g$ are pure twist-3 parton correlation functions
(PCF) defined in terms of the quark-gluon-quark correlator, which in
the light-cone gauge reads 
\begin{equation}
\begin{split}
&i \Phi_{F ij}^\alpha(x,x') \stackrel{\text{LC}}{=}
\int \frac{d \xi^- d \eta^-}{(2 \pi)^2}\,
e^{i k \cdot \xi}\, e^{i (k'-k) \cdot \eta}
\langle P| \bar{\psi}_j(0)\,
i g \, \partial^+_{\eta} A_T^{\alpha}(\eta) \, 
\psi_i(\xi) |P \rangle 
\Big|_{\substack{
\xi^+=\xi_T=0 \\
\eta^+=\eta_T=0}} \ ,
\label{e:phiF}
\end{split}  
\end{equation}
where $\alpha$ is a transverse index, 
$x' = \tfrac{k'\cdot n_-}{P\cdot n_-}$  and  $F$ is the QCD
field strength tensor. 
The Lorentz decomposition of $\Phi_F$ 
defines the relevant PCFs \cite{Boer:1997bw,Kanazawa:2000hz},
\begin{equation}
\begin{split} 
i\Phi_F^{\rho}(x,x')& = \frac{M}{4}\biggl[ G_F(x,x') i \epsilon_T^{\rho \alpha}S_{T
    \alpha} + \tilde{G}_F(x,x')S_T^{\rho}\gamma_5 + \ldots \biggr] \nslash_+\ ,
\end{split} 
\end{equation} 
where hermiticity and parity constrain $G_F(x,x') = G_F(x',x)$ and $ 
\tilde{G}_F(x,x') = -\tilde{G}_F(x',x)$. The pure twist-3 functions in
Eqs.~\eqref{eq:EOM}-\eqref{eq:LIR} are particular projections over
$x'$ of $G_F(x,x')$ and $\tilde G_F(x,x')$(PV denotes the principal value):
\begin{align}
  x \tilde{g}_T^a(x) &=  \text{PV} \int dx'\, 
    \frac{G_F(x,x') + \tilde{G}_F(x,x')}{2(x'-x)}.
    \label{e:tildeg} \\
  \hat{g}_T(x)^a & = \text{PV} 
    \int d x'\, \frac{\tilde{G}_F(x,x')/(x-x')}{x-x'},
    \label{e:hatg}
\end{align} 
and as such are sensitive to different parts of the quark-gluon-quark
correlator. 
It is very important to find several such quantities, because 
physically it is only possible to measure $x$ but the full 
dependence on $(x,x')$ is needed, e.g., to determine the QCD
evolution of $g_2$ or to compute the high-$k_T$ tails of TMDs. 
Note also that since the integrand in Eq.~\eqref{e:hatg} is
antisymmetric in $x$, $x'$, we obtain the non trivial property 
\begin{equation}
   \int_0^1 dx\, \hat{g}_T^a(x) = 0 \ .
\label{eq:hatsumrule}
\end{equation} 

\section{The WW relation}

Eliminating $g_{1T}^{a(1)}$ from \eqref{eq:EOM}-\eqref{eq:LIR} one can
derive the Wandzura-Wilczek relation \eqref{eq:WWrel} for the structure
function $g_2 = -g_1 + \frac12 \sum_a g_T^2$, and explicitly write
down its breaking term $\Delta = g_2 - g_2^{\rm{WW}}$:
\begin{equation} 
\begin{split} 
 g_2(x) - g_2^{\rm{WW}}(x) 
    = \frac{1}{2}\,\sum_a e_a^2
\biggl(
    \tilde g_T^a(x) - \int_{x}^1 \frac{dy}{y} \tilde g_T^a(y) 
    + \int_{x}^1\frac{dy}{y} \hat{g}_T^a(y) 
\Biggr)\ ,
\label{eq:WWrelation_mod}
\end{split} 
\end{equation} 
Note that $g_2$ explicitly satisfies the Burkhardt--Cottingham sum rule 
$\int_0^1 g_2(x) = 0$, which is not in general guaranteed in the
OPE~\cite{Jaffe:1989xx,Jaffe:1990qh}.

A natural question is: how much is the WW relation broken? 
Model calculations have been used to repeatedly argue 
that the pure twist-3 terms are not necessarily
small~\cite{Jaffe:1990qh,Harindranath:1997qn}. 
However, in the recent past, since the LIR-breaking $\hat{g}_T$ term
was not considered in 
Eq.~\eqref{eq:WWrelation_mod} and the quark-mass term with $h_1$ was neglected,
the breaking of the WW relation was considered to be a direct measurement of
the pure twist-3 term $\tilde{g}_T$. 
Therefore, the presumed experimental validity of the WW relation,
which we are presently going to challenge, was taken as
evidence that $\tilde{g}_T$ is small. This observation was also typically
generalized to assume that all pure twist-3 terms are small. 

\begin{figure}[bt]
  \centering
  \includegraphics[width=0.49\linewidth]{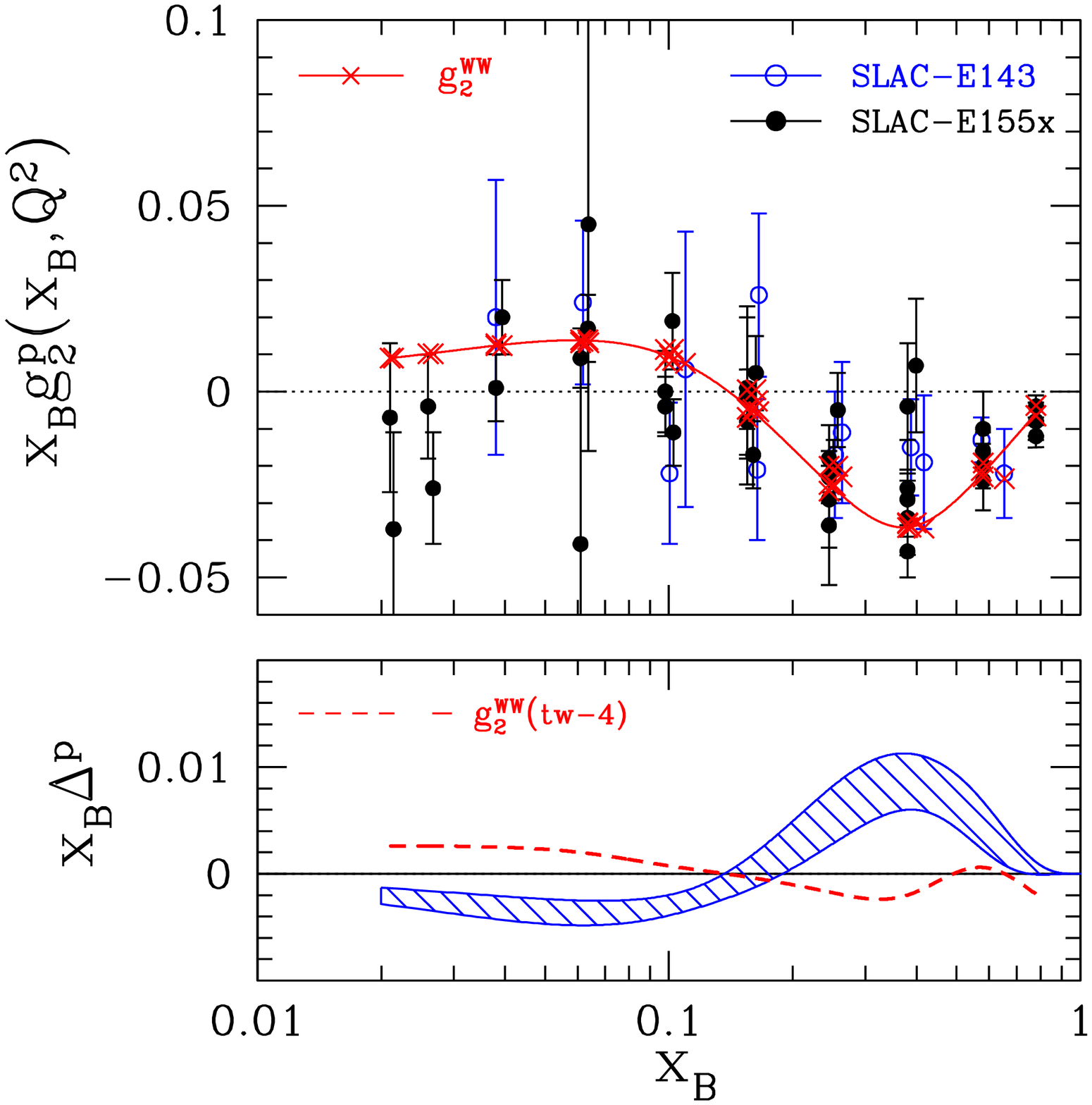}
  \includegraphics[width=0.49\linewidth]{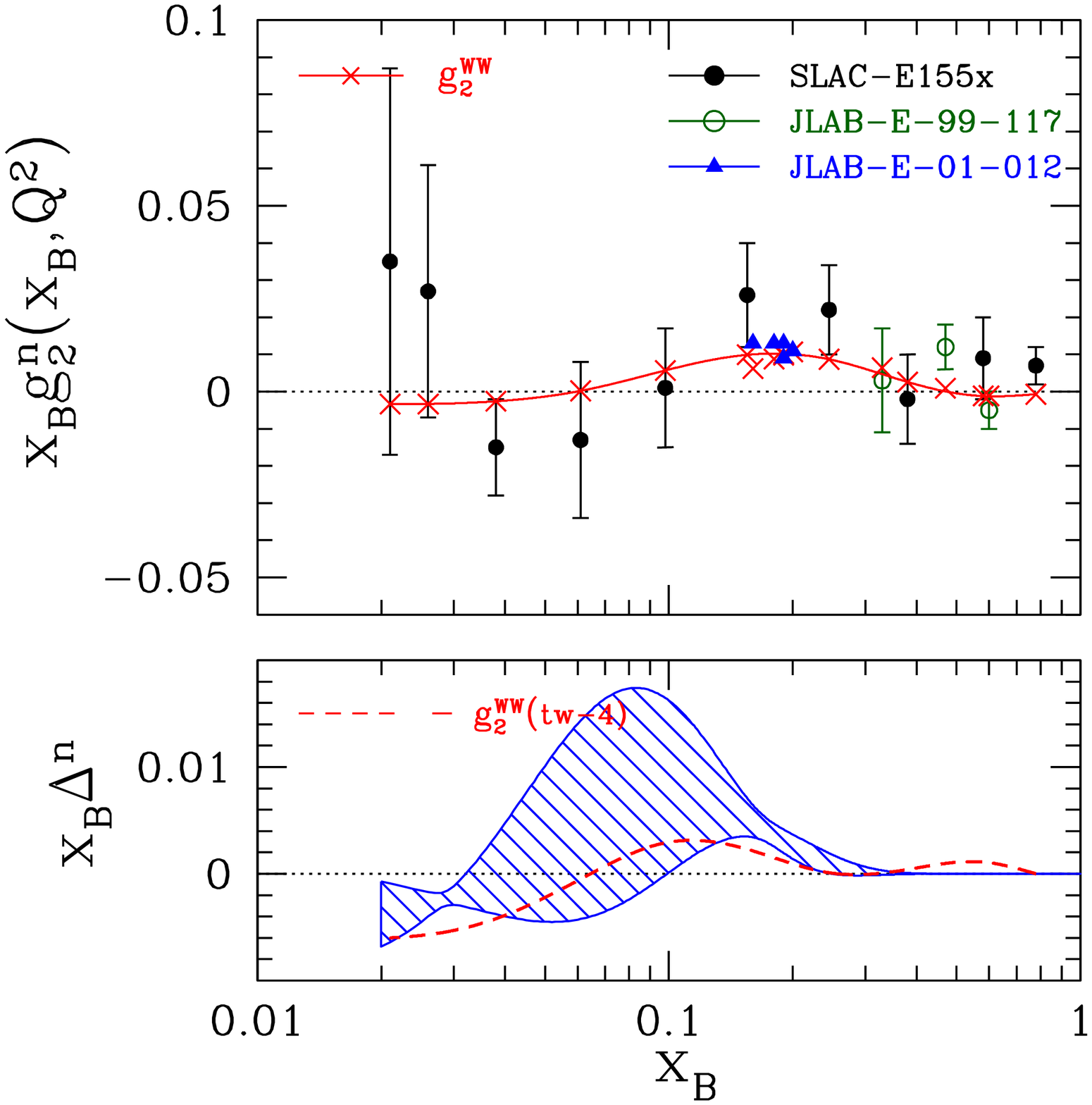}
  \caption{
    {\it Top panels}: 
    the experimental proton and neutron $g_2$ structure
    function compared to $g_2^{\rm{WW}}$. The crosses are $g_2^{\rm{WW}}$
    computed at the experimental kinematics. The lines are  $g_2^{\rm{WW}}$
    computed at the average $Q^2$ of the E155x experiment: the solid (dashed)
    line is computed with the LSS2006 fits of $g_1$, with (solid) and
    without (dashed) the HT contribution obtained in the fit.
    Data points for the proton target \cite{Abe:1998wq,Anthony:2002hy}
    have been slightly shifted in $x$ for clarity. For the neutron
    only the high precision data from
    \cite{Anthony:2002hy,Zheng:2004ce,Kramer:2005qe} have been 
    included. 
    {\it Bottom panels}:
    The WW-breaking term $\Delta_\text{th}$ for model (I) and (II)
    compared to the higher-twist contribution to $g_1$. See text for
    further details. 
  }
  \label{fig:g2WW}
\end{figure}

Our present analysis shows instead that, precisely due to the presence
of $\hat{g}_T$, the measurement of the breaking of the WW relation 
does not offer anymore the possibility of measuring a single pure
twist-3 matrix element, nor to generically infer its size. 
On the theory side, the quark-target model of
Refs.~\cite{Harindranath:1997qn,Kundu:2001pk} can be used to determine 
both $\tilde{g}_T$ and $\hat{g}_T$, which are both comparable in size to the 
the other twist-2 functions. On the experimental side, we used
data on polarized DIS on proton and neutron targets to
fit the WW breaking term $\Delta(x)$ defined as the
difference of the experimental data and $g_2^{\rm{WW}}$: 
\begin{align}
  \Delta(x) = g_2^{\rm{ex}}(x,Q^2) - g_2^{\rm{WW}}(x,Q^2) \ .
\end{align}
$g_2^\text{WW}$ was determined using the LSS06 leading twist $g_1$
parametrization \cite{Leader:2006xc}, and $\Delta$ fitted to
a functional form allowing for a change in sign and satisfying the
Burkhardt--Cottingham sum rule. 
The result is presented in Fig.~\ref{fig:g2WW}, and
Table~\ref{tab:WWfits}, where the deviation from the WW relation is 
quantified for a given $[x^{\rm{min}},x^{\rm{max}}]$ interval by 
\begin{align}
  r^2 = \frac{\int_{y^{\rm{min}}}^{y^{\rm{max}}} dy\, x\Delta_{\rm{th}}^2(x)}
           {\int_{y^{\rm{min}}}^{y^{\rm{max}}} dy\, xg_2^2(x)} \ ,
\end{align}
with $y=\log(x)$. The value of $r$ is a good approximation to
the relative magnitude of $\Delta$ and $g_2$, which are sign-changing
functions. For the proton, we considered three intervals: the whole
measured $x$ range, [0.02,1]; the low-$x$ region, [0.02,0.15]; the
large-$x$ region, [0.15,1]. For the neutron, due to the limited statistical
significance of the low-x data, we limit ourselves to quoting the value of $r$
for the large-$x$ region, [0.15,1]. 

In summary, we have found that the experimental data are compatible with a
substantial breaking of the WW relation in the 15-40\% range.

\begin{table}[bt]
\centering
\begin{tabular}[c]{crlcccc}
\hline & \multicolumn{2}{l}{\bf proton} & $\chi^2$/d.o.f. 
& $r_\text{tot}$ & $r_\text{low}$ & $r_\text{hi}$ \\\hline
  (I)  & $\Delta_{\rm{th}}$ & = 0 
    & 1.22 \hspace*{0cm} \\
  (II)  
    & $\Delta_{\rm{th}}$ & = $\alpha (1-x)^\beta \bigl((\beta+2) x-1\bigr)$
                                        \hspace*{.1cm}\\
    & $\alpha$ & = $0.13 \pm 0.05$ \\
    & $\beta$ & = $4.4 \pm 1.0$ 
    & 1.05  &  15-32\% & 18-36\% & 14-31\% \\
\hline & \multicolumn{2}{l}{\bf neutron} \\\hline
  (I)   & $\Delta_{\rm{th}}$ & = 0 
    & 1.66 \hspace*{0cm} \\
  (II)  
    & $\Delta_{\rm{th}}$ & = $\alpha (1-x)^\beta \bigl((\beta+2) x-1\bigr)$
                                        \hspace*{.1cm}\\
    & $\alpha$ & = $0.64 \pm 0.92$ \\
    & $\beta$ & = $24 \pm 10$ 
    & 1.11  &   &  & 18-40\% \\
\hline
\end{tabular}
\caption{Results of the 1-parameter fits of the WW breaking term
  $\Delta_{\rm{th}}$ for different choices of its functional form.
  value $r$ of the relative size of the breaking 
  term is computed for the whole measured $x$ range, [0.02,1]; the
  low-$x$ region, [0.02,0.15]; the large-$x$ region, [0.15,1]. 
  See text for further details.} 
\label{tab:WWfits}
\end{table}


\section{A proposal for an experimental campaign}

Figure~\ref{fig:g2WW} clearly shows the need for better precision in
$g_2$ measurements with both proton and neutron targets. In
particular, for the neutron high precision is needed away from
$x\approx 0.15-0.20$ where JLab E01-012 data almost completely
determine the presented fits.
But even if in the future 
the WW approximation is found to be more precise than in our analysis,
we would only be able to conclude that  
\begin{equation}
   \sum_a e_a^2
\biggl(
    \tilde g_T^a(x) - \int_{x}^1 \frac{dy}{y} \tilde g_T^a(y) 
    + \int_{x}^1\frac{dy}{y} \hat{g}_T^a(y) \biggr)
    \approx 0 \ .
\label{eq:WWexper}
\end{equation}
This can clearly happen because either $\hat{g}_T$ and $\tilde g_T$ are
both small, or because they accidentally cancel each
other. Therefore no information can be obtained on the size of the
twist-3 quark-gluon-quark term $\tilde g_T$ from the 
experimental data on $g_2$ alone.\footnote{Note that these results were
essentially already obtained in Ref.~\cite{Metz:2008ib}. 
In that work, however, the authors {\em assumed} 
$\tilde{g}_T$ small and the WW relation small, obtaining a small
$\hat{g}_T$, which is unjustified as we
have just discussed.} 

However, individually determining the size of $\hat{g}_T$ and $\tilde
g_T$ is very important to gather information on the $x$, $x'$
dependence of the quark-gluon-quark correlator. This can be
experimentally accomplished by using the EOM \eqref{eq:EOM}  and LIR
\eqref{eq:LIR} and measuring the $g_{1T}^{(1)}$ function, accessible
in semi-inclusive deep inelastic 
scattering with transversely polarized targets and longitudinally polarized
lepton beams (see, e.g., Ref.~\cite{Bacchetta:2006tn}):
\begin{align}\begin{split}
  \hat{g}_T^a(x) & = g_{T}^a(x) - g_{1}^a(x) - \frac{d}{dx}\,g_{1T}^{a(1)}(x) \\
  \tilde{g}_T^a(x) &=  g_T^a(x) - \frac{1}{x} g_{1T}^{a(1)}(x) \ .
\end{split}\end{align}
In TMD factorization, $g_{1T}^{(1)}$ is the
first transverse moment of a twist-2 TMD. Its experimental determination is
challenging because it requires measuring a double spin asymmetry in
semi-inclusive DIS up to rather large hadron transverse momentum.
Furthermore, in the LIR it appears differentiated in $x$, which
requires a rather fine $x$ binning.
Preliminary data from the E06-014 and SANE (E-07-003) experiments at Jefferson
Lab will soon be available, and will demonstrate the feasibility of
the proposed measurement of $\hat{g}_T$ and $\tilde g_T$.

This measurement is also very important because 
the EOM \eqref{eq:EOM}, LIR \eqref{eq:LIR} and WW relation breaking
\eqref{eq:WWrelation_mod} provide 3 independent measurement for 2
independent quantities. Verifying them will constitute a pretty
stringent test of TMD factorization and its connection to collinear
factorization.


\begin{theacknowledgments}
This work was supported by the DOE contract No. DE-AC05-06OR23177,
under which Jefferson Science Associates, LLC operates Jefferson Lab,
and NSF award No. 0653508.
\end{theacknowledgments}






\begin{thebibliography}{24}

\bibitem[Accardi and Qiu(2008)]{Accardi:2008ne}
A.~Accardi, and J.-W. Qiu, \emph{JHEP} \textbf{07}, 090 (2008).

\bibitem[Accardi and Melnitchouk(2008)]{Accardi:2008pc}
A.~Accardi, and W.~Melnitchouk, \emph{Phys. Lett.} \textbf{B670}, 114--118
  (2008).

\bibitem[Ellis et~al.(1983)]{Ellis:1982cd}
R.~K. Ellis, W.~Furmanski, and R.~Petronzio, \emph{Nucl. Phys.} \textbf{B212},
  29 (1983).

\bibitem[Melnitchouk et~al.(2005)]{Melnitchouk:2005zr}
W.~Melnitchouk, R.~Ent, and C.~Keppel, \emph{Phys. Rept.} \textbf{406},
  127--301 (2005).

\bibitem[Farrar and Jackson(1975)]{Farrar:1975yb}
G.~R. Farrar, and D.~R. Jackson, \emph{Phys. Rev. Lett.} \textbf{35}, 1416
  (1975).

\bibitem[Brodsky et~al.(1995)]{Brodsky:1994kg}
S.~J. Brodsky, M.~Burkardt, and I.~Schmidt, \emph{Nucl. Phys.} \textbf{B441},
  197--214 (1995).

\bibitem[Isgur(1999)]{Isgur:1998yb}
N.~Isgur, \emph{Phys. Rev.} \textbf{D59}, 034013 (1999).

\bibitem{Burkardt:2009}
M.~Burkardt, contribution to these proceedings.

\bibitem[Belitsky(1997)]{Belitsky:1997ay}
A.~V. Belitsky  (1997), {hep-ph/9703432}.

\bibitem[Ji et~al.(2006)]{Ji:2006ub}
X.~Ji, J.-W. Qiu, W.~Vogelsang, and F.~Yuan, \emph{Phys. Rev. Lett.}
  \textbf{97}, 082002 (2006).

\bibitem[Wandzura and Wilczek(1977)]{Wandzura:1977qf}
S.~Wandzura, and F.~Wilczek, \emph{Phys. Lett.} \textbf{B72}, 195 (1977).

\bibitem[Jaffe(1990)]{Jaffe:1989xx}
R.~L. Jaffe, \emph{Comments Nucl. Part. Phys.} \textbf{19}, 239 (1990).

\bibitem[Jaffe and Ji(1991)]{Jaffe:1990qh}
R.~L. Jaffe, and X.~Ji, \emph{Phys. Rev.} \textbf{D43}, 724--732 (1991).

\bibitem[Accardi et~al.(2009)]{Accardi:2009}
A.~Accardi, A.~Bacchetta, W.~Melnitchouk, and M.~Schlegel, in
  preparation.

\bibitem[Boer et~al.(1998)]{Boer:1997bw}
D.~Boer, P.~J. Mulders, and O.~V. Teryaev, \emph{Phys. Rev.} \textbf{D57},
  3057--3064 (1998).

\bibitem[Kanazawa and Koike(2000)]{Kanazawa:2000hz}
Y.~Kanazawa, and Y.~Koike, \emph{Phys. Lett.} \textbf{B478}, 121--126 (2000).

\bibitem[Harindranath and Zhang(1997)]{Harindranath:1997qn}
A.~Harindranath, and W.-M. Zhang, \emph{Phys. Lett.} \textbf{B408}, 347--356
  (1997).

\bibitem[Kundu and Metz(2002)]{Kundu:2001pk}
R.~Kundu, and A.~Metz, \emph{Phys. Rev.} \textbf{D65}, 014009 (2002).

\bibitem[Leader et~al.(2007)]{Leader:2006xc}
E.~Leader, A.~V. Sidorov, and D.~B. Stamenov, \emph{Phys. Rev.} \textbf{D75},
  074027 (2007).

\bibitem[Abe et~al.(1998)]{Abe:1998wq}
K.~Abe, et~al., \emph{Phys. Rev.} \textbf{D58}, 112003 (1998).

\bibitem[Anthony et~al.(2003)]{Anthony:2002hy}
P.~L. Anthony, et~al., \emph{Phys. Lett.} \textbf{B553}, 18--24 (2003).

\bibitem[Zheng et~al.(2004)]{Zheng:2004ce}
X.~Zheng, et~al., \emph{Phys. Rev.} \textbf{C70}, 065207 (2004).

\bibitem[Kramer et~al.(2005)]{Kramer:2005qe}
K.~Kramer, et~al., \emph{Phys. Rev. Lett.} \textbf{95}, 142002 (2005).

\bibitem[Metz et~al.(2008)]{Metz:2008ib}
A.~Metz, P.~Schweitzer, and T.~Teckentrup, {arXiv:0810.5212 [hep-ph]}.

\bibitem[Bacchetta et~al.(2007)]{Bacchetta:2006tn}
A.~Bacchetta, M.~Diehl, K.~Goeke, A.~Metz, P.~J. Mulders, and M.~Schlegel,
  \emph{JHEP} \textbf{02}, 093 (2007).

\end{thebibliography}

\end{document}